\documentclass{ifacconf}

\usepackage{amsmath,amssymb} 
\usepackage{mathtools}
\usepackage[utf8]{inputenc} 

\usepackage[pdftex]{graphicx}
\graphicspath{{plots/}}

\usepackage{color}

\newcommand{\set}[1]{\mathcal{#1}}
\newcommand{\op}[1]{\mathrm{#1}}

\newcommand{\Tref}{v_\op{ref}}

\newcommand{\Il}{\ell} 

\newcommand{\EV}{\mathbb{E}} 

\newcommand{\tin}{\mathtt{t}} 

\usepackage{natbib}

\usepackage{flushend}

\begin{document}
 
\begin{frontmatter}

\title{Efficient Dynamic Programming Solution to a Platoon Coordination Merge Problem With Stochastic Travel Times\thanksref{footnoteinfo}} 

\thanks[footnoteinfo]{This work was supported by the COMPANION EU project, the Knut and Alice Wallenberg Foundation, the Swedish Research Council, and the Swedish Strategic Research Foundation.}

\author[First]{Sebastian van de Hoef} 
\author[First]{Karl H. Johansson} 
\author[First]{Dimos V. Dimarogonas}

\address[First]{ACCESS Linnaeus Center and the School of Electrical Engineering, KTH Royal Institute of Technology, SE-100 44, Stockholm, Sweden (e-mail: {\tt \{shvdh, kallej, dimos\}@kth.se}).}

\begin{abstract}                
The problem of maximizing the probability of two trucks being coordinated to merge into a platoon on a highway is considered. Truck platooning is a promising technology that allows heavy vehicles to save fuel by driving with small automatically controlled inter-vehicle distances. In order to leverage the full potential of platooning, platoons can be formed dynamically en route by small adjustments to their speeds. However, in heavily used parts of the road network, travel times are subject to random disturbances originating from traffic, weather and other sources. We formulate this problem as a stochastic dynamic programming problem over a finite horizon, for which solutions can be computed using a backwards recursion. By exploiting the characteristics of the problem, we derive bounds on the set of states that have to be explored at every stage, which in turn reduces the complexity of computing the solution. Simulations suggest that the approach is applicable to realistic problem instances.
\end{abstract}

\begin{keyword}
Transportation, Platooning, Dynamic Programming, Uncertainty, Coordination
\end{keyword}

\end{frontmatter}

\section{Introduction}

Truck platooning is a promising technology that enables significant fuel savings for heavy vehicles. It leverages automatic control of inter-vehicle distances allowing for small longitudinal spacing between trucks without affecting safety. This reduces the air resistance of the trailing vehicles in the platoon effectively which translates into a reduction in fuel consumption. Other benefits include improved road utilization, increased safety, and decreased workload for the driver. 
Truck platooning has been successfully demonstrated by several vehicle manufacturers, e.g., (\cite{cyberphysicalTransport,konvoi_ref,dolphin_framework_conference}).

The efficient management of platoon formation is a crucial ingredient for leveraging platooning (\cite{whitepaper_platooning}). 
We propose to form platoons dynamically en route by slightly adjusting the speed of the vehicles. It has been shown that coordinating platooning centrally can improve the platooning rate and the system level reduction in fuel consumption significantly over spontaneous platooning where trucks form platoons if they happen to get into each others vicinity (\cite{ITSCpaper}).  

Platoon coordination has been approached on different levels of abstraction, including combination of platooning and routing (\cite{Larsson_Platoon_Complexity}), identification of promising platoon partners (\cite{datamining_platooning}), as part of automated highway systems (\cite{path_overview_conference}), and using local infrastructure based controllers (\cite{jeff_kuo_yun_distributed_controller}).
Our previous work (\cite{ITSCpaper}) proposes a framework in which pairwise plans are systematically composed into an overall coordination plan for all vehicles. The proposed planning assumes that the speed can be deterministically selected within a small range of feasible speeds. To this end, the upper bound on the speed on a road segment can be estimated by using historic data, traffic measurements and advanced prediction models
(\cite{traffic_state_estimation_papgeorgiou, traffic_state_estimation_celikoglu, traffic_state_estimation_horowitz}).
However, an accurate prediction of the travel times in a road network is a challenging task and even advanced prediction models leave some uncertainty.
There is a wide scope of models that provide in addition to the expected travel time its distribution, typically with the goal of quantifying the reliability of road infrastructure, for instance, 
\cite{travel_time_reliability2,travel_time_reliability3,estimation_from_probe_vehicles1,travel_time_reliability_flow,truck_travel_time_reliability}. 

In this paper, we consider a scenario where two vehicles should merge at the intersection of their routes. One of the vehicles has a fixed reference speed while the reference speed of the other vehicle can be adjusted, fitting in the framework of \cite{ITSCpaper}. 
The objective is to control the second vehicle to maximize the probability of both vehicles arriving at the intersection with a time difference less than a given threshold. At the same time, this probability is to be computed as an input to the higher planning layer that combines pairwise plans into a plan for all vehicles that are coordinated by the platoon service provider at a given point in time. The considered distances to the merge point are larger than in settings like \cite{bart_optimal_control_merge, on_ramp_coordination_survey} and references therein, where vehicle dynamics and potentially all vehicles in the control zone can be explicitly considered. \cite{kuo_yun_traffic_platoon_formation} have employed traffic flow theory in a scenario where one vehicle catches up to the other on the same road. 
The main contribution of the paper is to formulate the platoon coordination merge problem in the framework of stochastic dynamic programming, which yields a controller that maximizes and explicitly computes the probability of a successful merge. Furthermore, we derive how to bound the subsets of states that need to be explored which is a prerequisite to computing solutions. By allowing for a freely selectable error tolerance these bounds are further improved. The method is demonstrated in a simulation example.  

The outline of the remainder is as follows.
The problem is formally modeled in Section~\ref{sec:problem_formulation}. In Section~\ref{sec:optimal_speed_control}, we formulate the dynamic programming solution and show how computing solutions can be made tractable. 
Section~\ref{sec:simulations} discusses simulation examples demonstrating the effectiveness of the method. Section~\ref{sec:conclusions} concludes the paper and outlines future work.

\section{Problem Formulation}
\label{sec:problem_formulation}

\begin{figure}[t]
\begin{center}
 \includegraphics[width=\columnwidth]{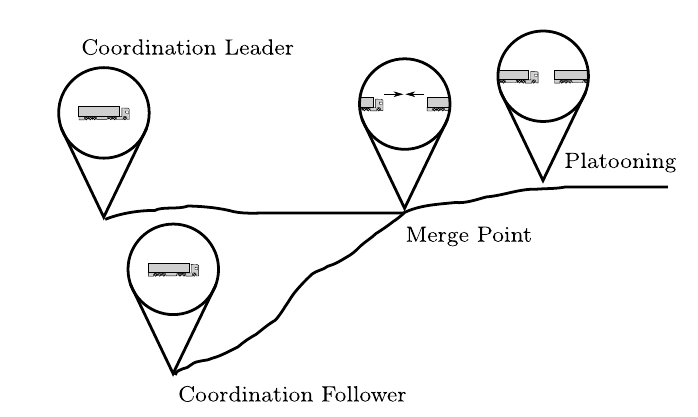}
 \caption{In the considered scenario, two vehicles are to merge into a platoon at the intersection of their routes by adapting their speed on the way leading to that intersection.}
\label{fig:scenario}
\end{center}
\end{figure}

Consider the scenario depicted in Fig.~\ref{fig:scenario} of two vehicles approaching an intersection at which they are supposed to merge into a platoon. One vehicle, the coordination leader, is controlled to arrive at the merge point at a specified point in time, and the other vehicle, the coordination follower, is controlled to maximize the benefit from platooning with the coordination leader. This is motivated by the framework introduced in \cite{ITSCpaper}, where several coordination followers are independently assigned to a coordination leader as the result of a discrete optimization problem. 

First, we model the movement of a single vehicle until the merge point. We consider that the route is partitioned into a finite number of segments. We consider discrete time and represent it as integers where the measurement unit is such that one increment corresponds to a sufficiently small discretization interval.
The traversal time $T^i$ of the $i$-th segment is a random variable. Let $t^i \in \mathbb{Z}$ be the time the vehicle starts traversing the $i$-th route segment, in the following referred to as segment arrival time.  

\begin{figure}[t]
\begin{center}
 \includegraphics[width=\columnwidth]{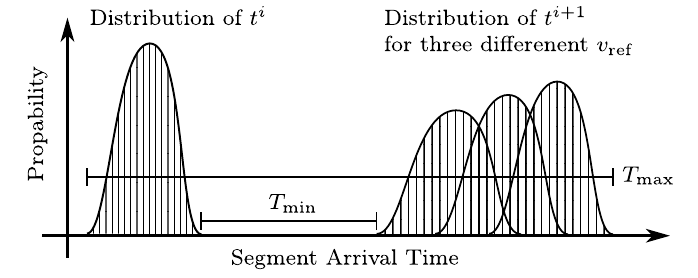}
 \caption{Illustration of how the distribution of $t^i$ and $t^{i+1}$ are related for different reference speeds $\Tref$.}
\label{fig:mdp_illustration}
\end{center}
\end{figure}

The arrival time at the next segment is the sum of the arrival time at the previous segment and the traversal time of the segment:
\begin{equation}
 t^{i+1} = t^{i} + T^i. \label{eq:arrival_time_recursion}
\end{equation}
The traversal time $T^i \in \mathbb{Z}$ is a random variable that is assumed only to be dependent on the reference speed at the $i$-th segment $\Tref^i \in \mathbb{V}$, which is considered to be a control input. The domain of $\mathbb{V}$ is a finite set of reference speeds. It is assumed that 
$
 T_{\min}^i \leq T^i \leq T_{\max}^i
$, see Fig.~\ref{fig:mdp_illustration}.

Since $T^i$ is assumed to depend only on the control input $\Tref^i$, eq.~\eqref{eq:arrival_time_recursion} describes a Markov decision process where $t^i$ denotes the value of its state at the $i$-th stage and $\mathbb{V}$ is the set of actions. Note that the stages in the decision process correspond to locations. Let $p_{T^i}(\tau|\Tref^i)$ denote the probability of $T^i = \tau$ conditioned on $\Tref^i$. The transition probability between state $t^{i}$ to $t^{i+1}$ is the probability that $T^i = t^{i+1} - t^{i}$, and thus the probability distribution of $t^i$ can be recursively computed as
\begin{equation}
\begin{split}
 p_{t^{i+1}}(\tin) &= \sum\limits_{\tau = -\infty}^{\infty} p_{T^i}(\tau | v_\op{ref}^i) p_{t^i}(\tin - \tau)\\
 &= \sum\limits_{\tau = T_{\min}^i}^{T_{\max}^i} p_{T^i}(\tau | v_\op{ref}^i) p_{t^i}(\tin - \tau). \label{eq:folding_recursion}
\end{split}
\end{equation}
Note that $p_{T^i}$ can also be modeled conditioned on the segment arrival time $t^i$ to reflect that travel time distributions are time dependent.
Let $t_\Il^i$ denote the segment arrival time of the coordination leader at the $i$-th segment of its route. We consider that the reference speed of the coordination leader is given as $v_\Il$ and its start time $t_\Il^\op{S}$ is known meaning that $p_{t_\Il^1}(\tin) = 1$ if $\tin = t_\Il^\op{S}$ and $p_{t_\Il^1}(\tin) = 0$ otherwise. Let $N_\Il$ be the index in the coordination leader's route at which the coordination leader and the coordination follower are supposed to meet. The probability distributions of $t_\Il^{N_\Il}$ are recursively computed from \eqref{eq:folding_recursion}.

We assume that a coordination leader and a coordination follower can platoon if they arrive at the merge point with an absolute time difference of at most $\Delta t$, which is chosen small enough so that they can establish vehicle-to-vehicle communication and initiate a merge maneuver.
 
The probability of platooning conditioned on that the arrival time of the coordination follower at the merge point $t_\op{f}^{N_\op{f}} = \tin$ is
\begin{equation*}
 P_\op{pl}(\tin) \coloneqq \mathrm{P}(|t_\Il^{N_\Il} - t_\op{f}^{N_\op{f}}| < \Delta t \;|\; t_\op{f}^{N_\op{f}} = \tin) =
 \sum\limits_{\tau = \tin - \Delta t}^{\tin + \Delta t} p_{t_\Il^{N_\Il}}(\tau).
\end{equation*}
The objective is to compute policies $v_\op{ref}^i:\mathbb{Z} \rightarrow \mathbb{V}$, $i = 1, \dots, N_\op{f}-1$ for selecting reference speeds 
for the coordination follower that maximize the expected value of $P_\op{pl}$ with respect the probability distribution of the coordination follower's arrival time at the merge point and conditioned on that the follower's start time at the first segment is $t_\op{f}^\op{S}$.
\begin{equation}
 \EV_{t_\op{f}^{N_\op{f}}}(P_\op{pl} | t_\op{f}^{1} = t_\op{f}^\op{S}) = \sum\limits_{\tin = -\infty}^{\infty} p_{t_\op{f}^{N_\op{f}}}(\tin|t_\op{f}^{1} = t_\op{f}^\op{S})   P_\op{pl}(\tin). \label{eq:objective}
\end{equation}
The relation between $t_\op{f}^{1}$ and $t_\op{f}^{N_\op{f}}$ is given by \eqref{eq:folding_recursion}.

\section{Optimal Speed Control}
\label{sec:optimal_speed_control}

In this section, we derive how to optimally select the reference speeds using dynamic programming. At the same time, the probability of platooning $\EV_{t_\op{f}^{N_\op{f}}}(P_\op{pl})$, as defined in \eqref{eq:objective}, is computed. This information can be used in order to decide whether or not two vehicles should platoon. More specifically, in the algorithm presented in \cite{ITSCpaper}, expected fuel savings from platooning would now be used instead of a predicted reduction in fuel consumption based on a deterministic model. 

The formulated problem fits the framework of optimal stochastic programming, when defining the value function as
$
 J^{i}(\tin) =  \EV_{t_\op{f}^{N_\op{f}}}(P_\op{pl} | t_\op{f}^{i} = \tin),
$
where $\EV_{t_\op{f}^{N_\op{f}}}(P_\op{pl} | t_\op{f}^{i} = \tin)$ is the expected value of $P_\op{pl}$ conditioned on that $t_\op{f}^{i} = \tin$ and under the optimal policies $v_\op{ref}^j$, $j = i, \dots, N_\op{f}-1$.

The value function
at the final stage is accordingly
\begin{equation}
J^{N_\op{f}}(\tin) =   \EV_{t_\op{f}^{N_\op{f}}}(P_\op{pl} | t_\op{f}^{N_\op{f}} = \tin) = P_\op{pl}(\tin), \label{eq:final_state_cost}
\end{equation}
and there is no stage cost.

The dynamic programming backwards recursion becomes 
\begin{equation}
 J^{i-1}(\tin) 
 = \max\limits_{v_\op{ref}^i} \left( \sum\limits_{\tau = -\infty}^{\infty} 
 p_{T^{i-1}_\op{f}}(\tau|v_\op{ref}^i)J^{i}(\tin+\tau) 
 \op{d} \tau \right). \label{eq:dynamic_programming_recursion}
\end{equation}
The probability of a successful merge when the coordination follower starts at time $t_\op{f}^\op{S}$ is given by
$
 J^1(t_\op{f}^\op{S}) = \EV_{t_\op{f}^{N_\op{f}}}(P_\op{pl} | t_\op{f}^{1} = t_\op{f}^\op{S}),
$
and the argument of the maximization in \eqref{eq:dynamic_programming_recursion} yields the optimal policy for each stage.

\subsection{Correlated Travel Time Distributions}

This section describes how the previously derived method can be extended to the case where travel times are correlated between segments. 
Depending on the length of the segment and the amount of factors $T^i$ is conditioned on for prediction, $T^i$ might be dependent on segments that are geographically close. 

We only consider correlation of travel times within the coordination leader's and the coordination follower's route.
In the framework of dynamic programming, the above reasoning implies that we have to add upstream traversal times $\mathbf{T}^i = [T^{i-1}, T^{i-2}, \dots, T^{i-H}]$ with horizon length $H \geq 1$ to the state,  which previously consisted only of the segment arrival time $t$, in order to retain the Markov property. The horizon length $H$ depends on the probabilistic model of the travel times. Equation~\eqref{eq:arrival_time_recursion} gets then augmented to:
\begin{equation*}
[
 t^{i+1},
 T^{(i-1)+1},
 \hdots,
 T^{(i-H)+1}
]
=
[
 t^{i} + T^i,
 T^{i},
 \hdots,
 T^{i-(H-1)}
].
\end{equation*}
A state $[\tin, \mathtt{T}^{-1}, \mathtt{T}^{-2}, \dots, \mathtt{T}^{-H}]$ can be reached from $[\tin-\mathtt{T}^{-1}, \mathtt{T}^{-2}, \mathtt{T}^{-3}, \dots, \mathtt{T}^{-(H-1)}, \tau]$ for any $\tau \in \mathbb{Z}$ with probability $p_{T^i}(\mathtt{T}^{-1} | \mathtt{T}^{-2}, \mathtt{T}^{-3}, \dots, \mathtt{T}^{-(H-1)}, \tau)$, so that
\eqref{eq:folding_recursion} becomes  
\begin{equation*}
\begin{split}
 &p_{(t^{i+1}, \mathbf{T}^{i+1})}(\tin, \mathtt{T}) =\\
 &\sum\limits_{\tau = T_{\min}^{i-H}}^{T_{\max}^{i-H}} 
 p_{T^i}(\mathtt{T}^{-1} | v_\op{ref}^i, \mathbf{T}^i = \mathtt{T}^{-}(\tau)) p_{(t^i, \mathbf{T}^i)}(\tin - \mathtt{T}^{-1}, \mathtt{T}^{-}(\tau)),
\end{split}
\end{equation*}
where $p_{(t^i, \mathbf{T}^i)}$ denotes the joint probability distribution of $t^i$ and $\mathbf{T}$, and where $\mathtt{T} = [\mathtt{T}^{-1}, \dots, \mathtt{T}^{-H}]$ and $\mathtt{T}^{-}(\tau) = [\mathtt{T}^{-2}, \dots, \mathtt{T}^{-(H-1)}, \tau]$. 

Note the difference in notation between the random variable $\mathbf{T}^i$ and a concrete value $\mathtt{T}$ that $\mathbf{T}^i$ can take. 
The terminal value is similar to \eqref{eq:final_state_cost}
$
J^{N_\op{f}}(\tin, \mathtt{T}) =  P_\op{pl}(\tin),
$
where $\mathtt{T} \in \mathbb{Z}^H$ and where the distribution $p_{t_\Il^{N_\Il}}$ is computed by marginalizing the traversal time states.

The backwards recursion as in \eqref{eq:dynamic_programming_recursion} changes to
\begin{equation*}
\begin{split}
 &J^{i-1}(\tin, \mathtt{T}) =\\ 
 &= \max\limits_{v_\op{ref}^i} \left( \sum\limits_{\tau=-\infty}^{\infty} 
 p_{T^{i-1}_\op{f}}(\tau|v_\op{ref}^i, \mathtt{T})J^{i}(\tin+\tau, \mathtt{T}^{+}(\tau)) 
 \op{d} \tau \right),
\end{split}
\end{equation*}
where $\mathtt{T}^{+}(\tau) = [\tau, \mathtt{T}^{-1}, \dots, \mathtt{T}^{-(H-1)}]$.

While it is straightforward to keep a record of previous segment traversal times, measuring traversal times of segments before the start is not trivial. If we assume that the computation of the platooning probability and the policies happen shortly before the vehicles start driving, real-time information from other sources such as traffic sensors or other vehicles might be used. Otherwise, travel time distributions that are not conditioned on segments before the start of the route have to be used. 

It is well known that the complexity of dynamic programming increases exponentially with the size of the state, an effect known as the curse of dimensionality. However, if we assume that most correlation between segment traversal time is actually caused by traffic dynamics (\cite{travel_time_reliability_flow}), we can reduce the state space size. According to macroscopic traffic flow theory, traffic has mainly three states~(\cite{three_phase_traffic}): free flow, synchronized flow, and congested flow. Therefore, the elements of $\mathbf{T}^i$ could potentially be discretized into these three regimes.

\subsection{Efficient Computation of Optimal Control Policies}

A challenge in using dynamic programming is finding ways of implementing the recursion and handling its complexity. There are two features make the problem considered in this paper computationally tractable. The first is the finite horizon of the problem and the second is the low dimensionality of the state space. Additionally, we can exploit the fact that unlike many other control systems, the objective of letting the trucks meet at the designated merge point does not have to be achieved at all cost. In case the merge fails, the problem can be resolved on the higher planning layer. Because of this property, it is reasonable to omit exploring state trajectories that lead to a successful merge but have low probability.  
 
We show that $J^i$ only has to be computed for an interval $[\underline{t}^i, \bar{t}^i]$ if a small error $\epsilon \geq 0$ on the computation of $J$ can be accepted. Furthermore, the length of the interval, i.e., $\bar{t}^i - \underline{t}^i$ does not depend on the stage $i$. We define $\underline{t}^i$ as
\begin{equation*}
 \underline{t}^{i+1} = \underline{t}^{i} + T_{\min}^i \Rightarrow \underline{t}^i = t_\op{f}^1 + \sum\limits_{j = 1}^{i-1} T_{\min}^i,
\end{equation*}
with $\underline{t}^{1} = t_\op{f}^\op{S}$, with $t_\op{f}^\op{S}$ being the start time of the coordination follower. 
Similarly, we define $\bar{t}^i$ as
\begin{equation*}
 \bar{t}^{i} = \bar{t}^{i+1} - T_{\min}^i \Rightarrow \bar{t}^{i} = \bar{t}^{N_\op{f}} - \sum\limits_{j=i}^{N_\op{f}-1} T_{\min}^i,
\end{equation*}
where $\bar{t}^{N_\op{f}}$ is selected large enough so that
\begin{equation}
 J^{N_\op{f}}(t) \leq \epsilon\text { for } t > \bar{t}^{N_\op{f}}, \label{eq:t_upper_end_def}
\end{equation}
for a given error tolerance $\epsilon \geq 0$.

Furthermore, we define an approximation of $J^i$ denoted as $\tilde{J}^i$. It is initialized at $i = N_\op{f}$ by
\begin{equation}
 \tilde{J}^{N_\op{f}}(\tin) = \left\{
 \begin{array}{ll}
  J^{N_\op{f}}(t) &\text{ if } \tin \in [\underline{t}^{N_\op{f}}, \bar{t}^{N_\op{f}}]\\
  0 &\text{ if } \tin \notin [\underline{t}^{N_\op{f}}, \bar{t}^{N_\op{f}}],
 \end{array}
 \right. \label{eq:J_tilde_end_def}
\end{equation}
and analogously to \eqref{eq:dynamic_programming_recursion} for $\tin \in [\underline{t}^{i-1}, \bar{t}^{i-1}]$
\begin{equation*}
 \tilde{J}^{i-1}(\tin) = 
          \max\limits_{v_\op{ref}^i} \left( \sum\limits_{\tau = \underline{t}^i}^{\bar{t}^i} 
	  p_{T^{i-1}_\op{f}}(\tau - \tin|v_\op{ref}^i)\tilde{J}^{i}(\tau) 
	  \right)
\end{equation*}
where the summation is rewritten in terms of the arrival time rather than the traversal time.

Note that segment arrival times $t^i \leq \underline{t}^i$ cannot be reached from $t_\op{f}^1 = t_\op{f}^\op{S}$, and therefore $J^i$ and likewise $\tilde{J}^i$ do not need to be computed for these times. Furthermore, $\tilde{J}^i(\tin) = 0$ for $i = 1,\dots,N_\op{f}$ and $\tin > \bar{t}^i$. 
The following result on the error between $J^i$ and $\tilde{J}^i$ holds
\begin{prop}
 For given error tolerance $\epsilon \geq 0$ it holds that
\begin{equation*}
 0 \leq J^{i}(\tin) - \tilde{J}^{i}(\tin) \leq \epsilon,
\end{equation*}
for all $i = 1, \dots, N_\op{f}$ and $\tin \geq \underline{t}^i$.
\end{prop}
The proof is omitted due to space constraints.

This proposition states that we underestimate the probability of platooning by at most $\epsilon$ when using $\tilde{J}^i$ instead of $J^i$. Choosing $\epsilon$ large means that $\bar{t}^i - \underline{t}^i$ is small which translates into small computational complexity but larger errors on the computation of $J^i$ and vice versa, since $\tilde{J}^i$ only needs to be computed in the interval $[\underline{t}^i, \bar{t}^i]$.
For $\tin > \bar{t}^i$, the chance of platooning with the coordination leader is smaller than $\epsilon$. Once the vehicle reaches a segment later than $\bar{\tin}^i$, it would no longer try to platoon with this coordination leader and instead either try to join another coordination leader or drive alone.
The computational complexity can potentially be significantly reduced by this approach depending on how spread the distribution of $T_\Il^{N_\Il}$ is. In practice, we would expect that $T_\Il^{N_\Il}$ with high variance also leads to platooning probabilities so small that they can in any case be discarded regardless of the follower's arrival time at the merge point.

\section{Simulations}
\label{sec:simulations}

Simulations are presented in this section in order to demonstrate the applicability of the derived results. The model published in \cite{truck_travel_time_reliability} is adapted for modeling the traversal time distributions. In \cite{truck_travel_time_reliability}, speed distributions are modeled as the mixture of Gaussian distributions, i.e., $p_V(v)  = W \set{N}(v, \mu_1, \sigma_1) + (1 - W)\set{N}(v, \mu_2, \sigma_2)$. We interpret one mode as corresponding to the free-flow and one to the congestion regime between which transition often happens suddenly (\cite{three_phase_traffic}). 

We model the effect of control by setting the mean value of the free flow model $\mu_2$ to $v_\op{ref}^i$, and we consider reference speeds in the range $v_\op{ref}^i \in [70\,\op{km}/\op{h}, 90\,\op{km}/\op{h}]$ with increments of 1\,km/h. Furthermore, both Gaussians are truncated individually to the range of $[10\,\op{km}/\op{h}, 100\,\op{km}/\op{h}]$. The measured speed value distributions in \cite{truck_travel_time_reliability} contain some entries well above the truck speed limit of $60\,\op{mph} \approx 96.56\,\op{km}/\op{h}$. Limiting the maximum speed is justified considering that a centralized planning system would not recommend speeds above the legal speed limits.
The speed is assumed to be constant over a segment, i.e., we have that the probability $P(T^i \leq \tau) = P(V \geq L^i/\tau)$, where $L^i$ is the length of the $i$-th segment. The remaining parameters of the speed distribution are listed in Table~\ref{tab:distribution}. The first set of parameters corresponds to a reliable segment with high speeds and little variation in the speed. The other set of parameters corresponds to an unreliable segment with a high risk of small speeds due to congestion and a wide spread in possible speeds.

\begin{table}
\begin{center}
 \begin{tabular}{lll}
 Variable & Reliable & Unreliable \\\hline
 $W$ & 0.04 & 0.55\\
 $\mu_1$ & 64.45\,km/h& 38.64\,km/h\\
 $\sigma_1$ & 34.76\,km/h& 18.96\,km/h\\
 $\sigma_2$ & 8.22\,km/h& 9.96\,km/h
\end{tabular}
\caption{Parameters of the speed distributions.}
\label{tab:distribution}
\end{center}
\end{table}

First, the following scenario is considered. The coordination leader's route consists of three segments with length 4\,km, 4\,km, 5\,km respectively, and the leaders start at time $t_\Il^1 = 0$. The reference speed of the leader is $80$\,km/h. The coordination follower's route also consists of three segments with lengths $6, 4, 5$\,km, and the start time is computed so that the coordination leader and follower would meet if they kept a constant speed of 80\,km/h. All segments are considered to be of the reliable type.
The maximum tolerable error $\epsilon$ between $J$ and $\tilde{J}$ is set to 1\,\%, and the maximum time-gap for platooning $\Delta t$ is 0.01\,h = 36\,seconds. 
A time step corresponds $10^{-4}\,\op{h} = 0.36$\,s.
The method was implemented using CPython 2.7 with Numpy and Scipy and this example takes less than 50 milliseconds to compute on a Core i3 processor using only one core. 

\begin{figure}[t]
\begin{center}
 \includegraphics[width=\columnwidth]{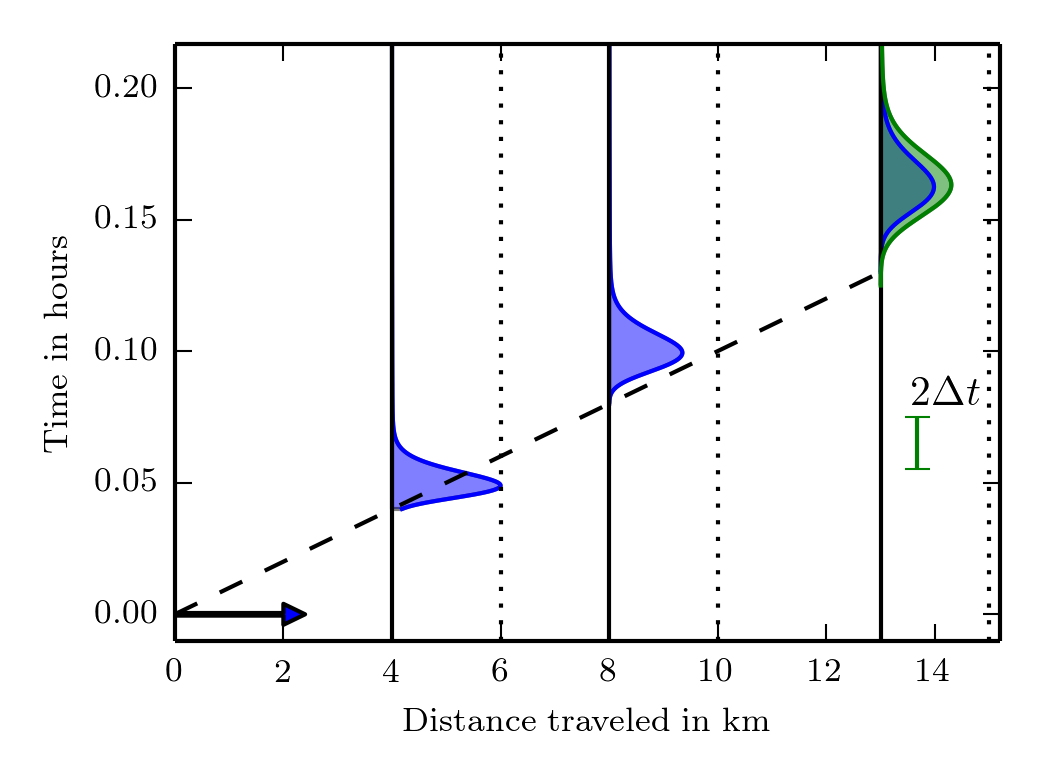}
 \caption{This plot shows the segment arrival time probability density functions of the coordination leader. At the beginning of the first segment, the start time is known and indicated by an arrow. The remaining densities are plotted on the vertical axis and are jointly scaled for presentation. The vertical dotted lines correspond to the maximum value of the first distribution and provide a reference for comparison of the distributions. The dashed line corresponds to the maximum speed of 100\,km/h. At the end of the last segment, which is the merge point, also $J^{N_\op{f}}$ is plotted in green. It is scaled so that a value of 1 corresponds to the level of the dotted line. The scale of $2 \Delta t$ is indicated on the right side of the plot. }
\label{fig:P_leader_over_distance}
\end{center}
\end{figure}

\begin{figure}[t]
\begin{center}
 \includegraphics[width=\columnwidth]{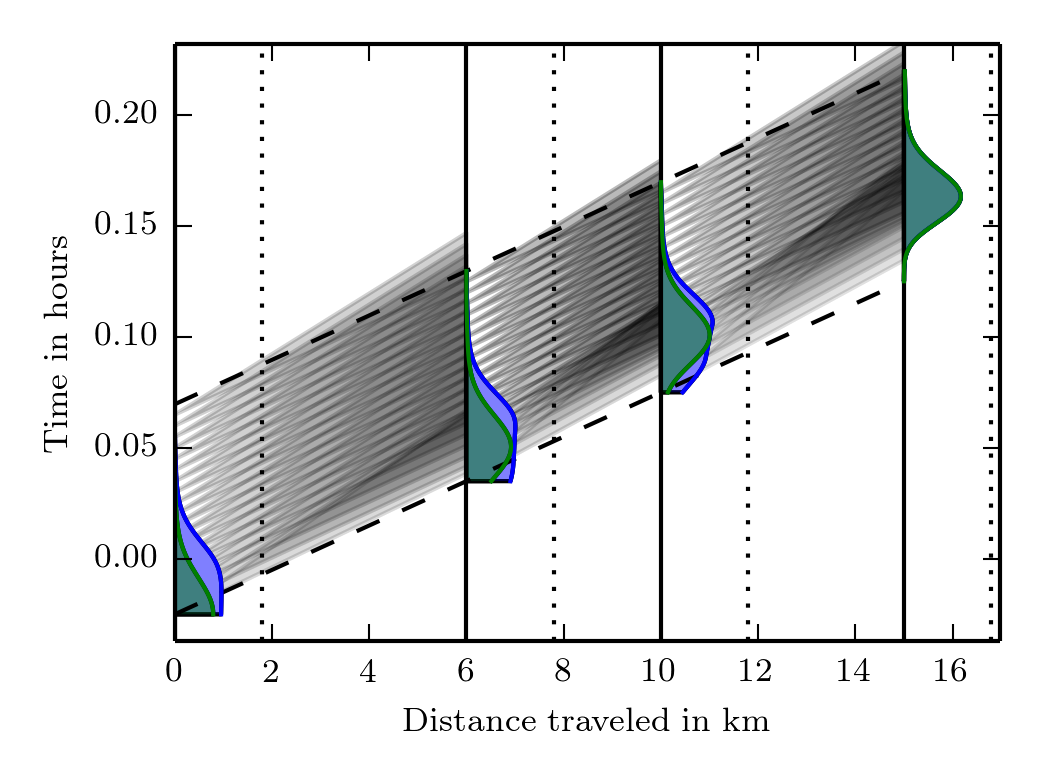}
 \caption{This plot shows $\tilde{J}^i$ at the distances from the coordination follower's start point corresponding to $i = 1,\dots,N_\op{f}$ as a function of segment arrival time. The vertical dotted lines correspond to a level of 1 one the scale of $\tilde{J}^i$. The blue plots show $\tilde{J}^i$ when the optimal control policy is implemented and the green curves when the reference speed is kept constant. The gray semi-transparent triangles visualize the control policy. The two rightmost corners of the triangle correspond to the 5- and 95 percentiles of the arrival times at a segment under the optimal control policy conditioned on that the start time at the previous segment corresponds to the leftmost corner of the triangle. The dashed lines correspond to $\underline{t}^i$ and $\bar{t}^i$.}
\label{fig:J_over_distance}
\end{center}
\end{figure}

Figs.~\ref{fig:P_leader_over_distance} and \ref{fig:J_over_distance} show the results from this scenario. Fig.~\ref{fig:P_leader_over_distance} shows the computed distributions of $T_\Il^i$. We can see that the distribution of $T_\Il^i$ gets spread out from segment to segment.  Fig.~\ref{fig:J_over_distance} shows the computed function $\tilde{J}^i$ with and without optimal control as well as a visualization of the optimal control policies. The optimal control is able to significantly improve the probability of platooning by centering the arrival time distributions of the next link at arrival times with high values for $\tilde{J}^i$. As $\tilde{J}^i$ gets more spread out to the left, the transition from slow reference speeds to the highest reference speeds with increasing segment arrival times also becomes more spread out. The start time has been chosen here in a way that the two trucks can easily meet in their reference speed range. This means that being unable to arrive sufficiently late is no issue and nothing could be gained from starting later. It is also possible to see how the optimal control is able to compensate if the coordination follower deviates from the trajectory that would be obtained by driving constantly at 80\,km/h. The merge probability $\tilde{J}^0(t_\op{f}^0)$ equals $52.96$\,\% using the optimal control and $43.97$\,\% using the fixed reference speed. The interval $\bar{t}_i - \underline{t}_i$ is $11.51$ times smaller than with $\epsilon = 0$, while the actual error on the platooning probability according to \eqref{eq:objective} from using $\tilde{J}^i$ instead of $J^i$ is $0.03\,\% \ll \epsilon = 1\,\%$.

\begin{figure}[t]
\begin{center}
 \includegraphics[width=\columnwidth]{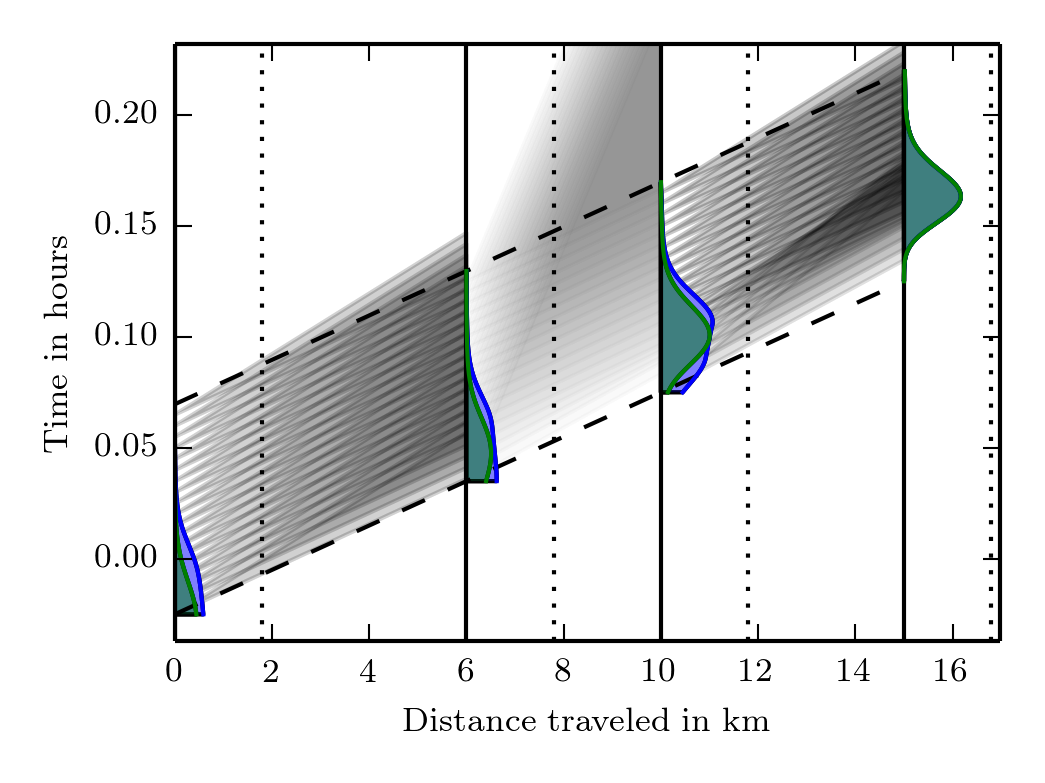}
 \caption{This plot shows, similar to Fig.~\ref{fig:J_over_distance}, $\tilde{J}^i$ at the distances from the coordination follower's start point corresponding to $i = 1,\dots,N_\op{f}$ as function of segment arrival time, but for a scenario where the second segment is unreliable.}
\label{fig:J_over_distance_congested}
\end{center}
\end{figure}

Fig.~\ref{fig:J_over_distance_congested} shows $\tilde{J}^i$ for a similar scenario as described above with the difference that the second segment in the coordination follower's route is unreliable. We can see that this causes a high risk of delay and thus much smaller values for $\tilde{J}^i$. Furthermore, the control policy selects higher reference speeds on the first segment compared to the previous scenario without an unreliable segment in order to compensate for a potential delay on the second segment. The merge probability $\tilde{J}^0(t_\op{f}^0)$ equals $32.62$\,\% using the optimal control and $24.62$\,\% using the fixed reference speed.

The two speed distributions are extreme cases of reliable and unreliable segments. In reality, there is probably a whole range of characteristics between these two extremes. Furthermore, it is unclear how a controlled truck would behave with respect to traversal time distributions, but we can assume that the variability would probably be smaller. A truck that is controlled to follow a reference speed behaves more predictably as the control reduces the variability due to different driver characteristics. In addition, we have to take into account that the spot speed can vary more than the traversal time. Take for instance a stop-and-go situation with regular shock-waves traveling upstream. In this case vehicles will exhibit a large variability in the speed which average out over a longer distance.
Nevertheless, the simulations demonstrate that the meth\-od can handle realistically sized instances of the problem and smaller variability would lead to even smaller intervals of $\bar{t}^i - \underline{t}^i$ and thus faster computation times. Note also that these kind of computations lend themselves well to parallel processing, for instance, on graphics cards.

\section{Conclusions}
\label{sec:conclusions}

The problem of maximizing the meeting probability of two vehicles at an intersection to form a platoon was formulated. In this model, control enters the system by affecting the traversal time distribution on segments leading towards the meeting point. 
The control problem was solved by means of dynamic programming which also yields the meeting probability explicitly as an input to higher planning layers. Considering the inherent constraint of the maximum speed of a truck and allowing for a small error makes it possible to significantly reduce the computational effort by effectively bounding the state-space in which solutions have to be computed. Simulations demonstrate the effectiveness of this approach.

In the future, we would like to test the algorithm with travel time distributions based on real truck travel time data, and analyze what the implications of integrating this kind of coordination in a large scale platoon coordination system are. Furthermore, it would be possible to consider a stage cost taking the dependency of fuel consumption on the speed into account. 
Additional optimizations on the numerical computation might be feasible in terms of adaptive time steps.

\end{document}